\documentclass[twocolumn,english,revtex,amsmath,amssymb,showpacs,float,longbibliography,superscriptaddress,nofootinbib]{revtex4-1}
\usepackage[T1]{fontenc}
\usepackage[latin9]{inputenc}
\usepackage{color}
\usepackage{babel}
\usepackage{amsmath}
\usepackage{stmaryrd}
\usepackage{graphicx}
\usepackage[unicode=true,
 bookmarks=false,
 breaklinks=false,pdfborder={0 0 1},backref=false,colorlinks=true]
 {hyperref}
\hypersetup{
 linkcolor=blue,citecolor=blue,urlcolor=blue,linkcolor=blue,citecolor=blue,urlcolor=blue}

\makeatletter

\usepackage{babel}

\makeatother

\begin{document}

\title{Extended Josephson junction qubit system}

\author{Andrey Grankin}

\author{Alicia J. Koll\'ar}

\author{Mohammad Hafezi}

\affiliation{Joint Quantum Institute, Department of Physics, University of Maryland,
College Park, MD 20742, USA}
\begin{abstract}
Circuit quantum electrodynamics (QED) has emerged as a promising platform
for implementing quantum computation and simulation. Typically, junctions
in these systems are of a sufficiently small size, such that only
the lowest plasma oscillation is relevant. The interplay between the
Josephson effect and charging energy renders this mode nonlinear,
forming the basis of a qubit. In this work, we introduce a novel QED
architecture based on extended Josephson Junctions (JJs), which possess
a non-negligible spatial extent. We present a comprehensive microscopic
analysis and demonstrate that each extended junction can host multiple
nonlinear plasmon modes, effectively functioning as a multi-qubit
interacting system, in contrast to conventional JJs. Furthermore,
the phase modes exhibit distinct spatial profiles, enabling individual
addressing through frequency-momentum selective coupling to photons.
Our platform has potential applications in quantum computation, specifically
in implementing single- and two-qubit gates within a single junction.
We also investigate a setup comprising several driven extended junctions
interacting via a multimode electromagnetic waveguide. This configuration
serves as a powerful platform for simulating the generalized Bose-Hubbard
model, as the photon-mediated coupling between junctions can create
a lattice in both real and synthetic dimensions. This allows for the
exploration of novel quantum phenomena, such as topological phases
of interacting many-body systems. 
\end{abstract}
\maketitle

\section{Introduction}

Josephson qubits, characterized by strong nonlinearity and low dissipation,
serve as promising building blocks for quantum computers \citep{devoret2013superconducting}
and quantum simulators of complex many-body systems \citep{HTK12,carusotto2020photonic}.
Fundamentally, these qubits encode quantum information in the form
of plasma oscillations occurring between two superconducting elements.
Although this is an intrinsically collective phenomenon within a many-electron
system, each junction typically operates within a regime that hosts
only a single qubit \citep{KSB20}. Various methods exist for interfacing
these qubits, such as coupling to a waveguide or an electromagnetic
resonator \citep{MCG07}. When the junction size is sufficiently small,
it can be treated as a lumped element, and its interaction with the
electromagnetic field is considered as an oscillating point-like dipole
\citep{BHW04,BGG21}.

However, more complex scenarios involving multiple junctions or a
multimode resonator present a challenging theoretical problem. These
are often addressed phenomenologically using the so-called black-box
quantization or energy-participation ratio approaches \citep{NPV12,minev2021energy}.
These methods are only valid in the weak nonlinearity limit, and the regime
of strong light-matter interaction is inaccessible. Furthermore, as the number of parameters increases, determining all coefficients within
the black-box quantization framework becomes impractical. More generally,
the dipole picture loses its validity when the junction size becomes
comparable to or larger than the Josephson penetration length ($\lambda_{J}$),
which characterizes the stiffness of phase fluctuations in junctions.
As a result, the extended quasi-1D junctions can host multiple plasmon
modes with the discrete spectrum given by:

\begin{equation}
\omega_{m}=\omega_{{\rm {pl}}}\sqrt{1+\left(\frac{\lambda_{J}}{L}\pi m\right)^{2},}\label{eq:Wm}
\end{equation}
where $\omega_{{\rm {pl}}}$ is the fundamental plasma frequency.
We note that in circuit-QED architecture, conventional qubits correspond
to the $m=0$ plasmonic mode as described in Eq.~\eqref{eq:Wm}.
This mode prevails in small junctions where $L$ is
significantly smaller than $\lambda_{J}$. On the other hand, as
we approach the opposite limit, the plasmon spectrum undergoes a transformation
into a continuous form \citep{T04,SSN08}.  This is accompanied by a reduction in the non-linearity of these modes, and
the full dynamics can be effectively described by the classical sine-Gordon
equations. While the investigation of this classical regime has a rich history \citep{UKC93,U98,KWU02}, given the technological developments in the past several decades, it is interesting to explore the quantum regime of such systems. The intermediate regime,  $L\sim\lambda_{J}$, is particularly interesting since one can expect several low-energy plasmon modes to be accessible while the strong nonlinearity is not compromised.

In this work, we investigate the light-matter interaction within extended
junctions, taking into consideration the inherent complexity of each
junction and its multiple degrees of freedom. From a conceptual point
of view, our setup could be considered a ``beyond dipole-approximation
circuit-QED architecture.\textquotedblright{} Besides, in contrast
to the black-box quantization, we follow a microscopic approach and
derive both linear and nonlinear terms of the corresponding Hamiltonian.
To this end, we develop a general theoretical framework for the quantum
description of extended junctions and their interaction with the electromagnetic
field. More precisely, we consider a fully microscopic model of two
infinitely thin superconducting layers and their coupling with the
resonator, which is depicted in Fig.~\ref{Fig1}~(a). We also microscopically
derive the Kerr and cross-Kerr non-linear interaction terms between
different plasmon modes. Effectively, each junction operates as an
interacting multi-qubit system, and we put forward a strategy for
performing single- and two-qubit gates. Furthermore, the spatial extent
of the electronic wavefunction for each plasmon mode in light-matter
coupling is leveraged to achieve complete qubit addressability, underscoring
its practicality for quantum computing applications.

We note that our approach allows us to not only to derive the black-box
quantization result in a \emph{bottom-up} way but also provides
a way to engineer effective Hamiltonians based on the mutual spatial
structure of the plasmon and resonator modes. In particular, we extend
our analysis to the case of many extended junctions coupled to a single
waveguide and demonstrate that such a setup could be used in order
to simulate complex many-body interacting Hamiltonians from generalized
Bose-Hubbard model to potentially lattice gauge theories. We note
that our architecture is distinct from the multi-junction circuit-QED
\citep{FSL17}, where the qubits do not interact unless they are coupled
via resonator. In contrast, in our proposal, nonlinear plasmon qubits
do interact with each other and their coupling can be controlled by
driving the system, i.e., a 1D array of them makes a 2D synthetic array
with full controllability. Our work is also distinct from the \emph{giant atom} scheme, where the JJ phase is a single number \citep{kockum2014designing,WLK21,kannan2020waveguide}. 

This paper is structured as follows. In Sec.~\ref{sec:Long-Josephson-junction},
we introduce classical sine-Gordon Lagrangian and perform quantization
of the phase fluctuations ignoring the nonlinearity. The latter is
then added perturbatively, inducing Kerr, cross-Kerr, and parametric
interaction terms between plasmon modes. This action can be obtained
from fully microscopic considerations as we show in Appendix ~\ref{sec:Microscopic-derivation}.
In Sec.~\ref{subsec:Coupling-to-the}, we derive the coupling of
the quantized fluctuations of the junction to the electromagnetic
field of a coplanar waveguide. In Sec.~\ref{sec:Quantum-simulation-of}-\ref{sec:qubit},
we propose an architecture for quantum computation and simulation
based on an array of extended Josephson junctions coupled to a single
waveguide.

\onecolumngrid 
\begin{center}
\begin{figure}[h]
\begin{centering}
\includegraphics[scale=0.35]{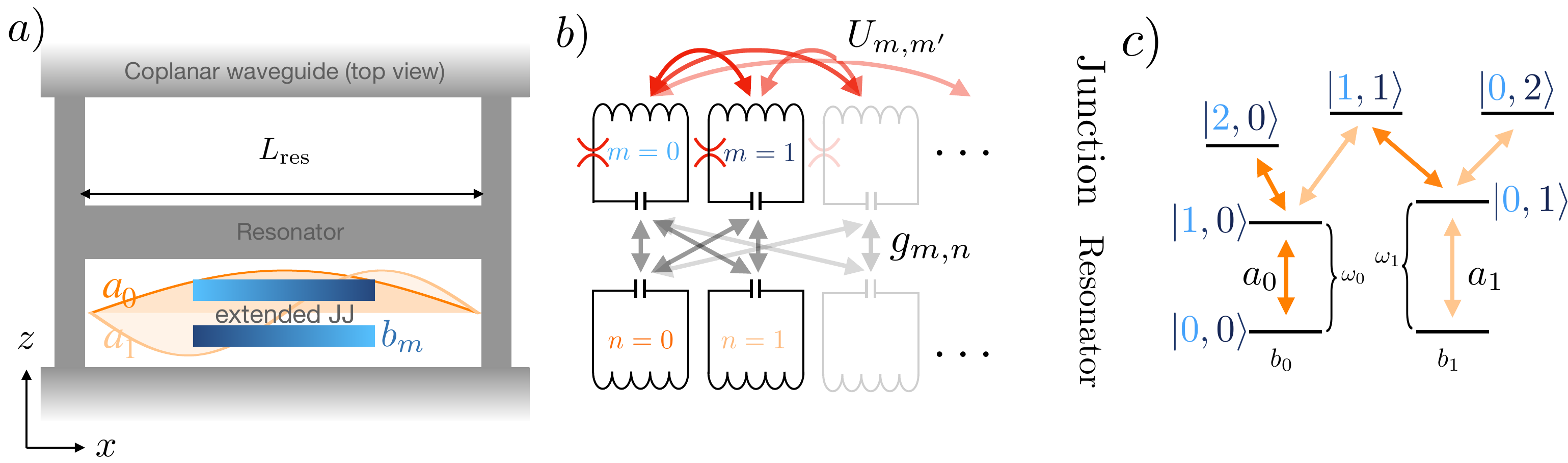} 
\par\end{centering}
\caption{Extended Josephson Junction qubit. (a) An extended Josephson junction
is placed in the center of the resonator. Superconducting phase fluctuations
modes $\left\{ b_{m}\right\} $ couple to the quantized resonator
modes $\left\{ a_{n}\right\} $. (b) Effective description of the
resonator-junction system in terms of equivalent LC circuits. Red
arrows and red spider symbols respectively indicate the cross- and
self-Kerr interaction terms. Grey arrows indicate the capacitative
coupling between junction and cavity modes. (c) Schematic representation
of the nonlinear two-qubit level scheme of a junction restricted to
the two lowest modes. Orange arrows indicate coupling to the resonator
modes due to the parity selection rules in case the junction is located
in the center of the resonator. }

\label{Fig1} 
\end{figure}
\par\end{center}

\begin{figure}[h]
\begin{centering}
\includegraphics[scale=0.3]{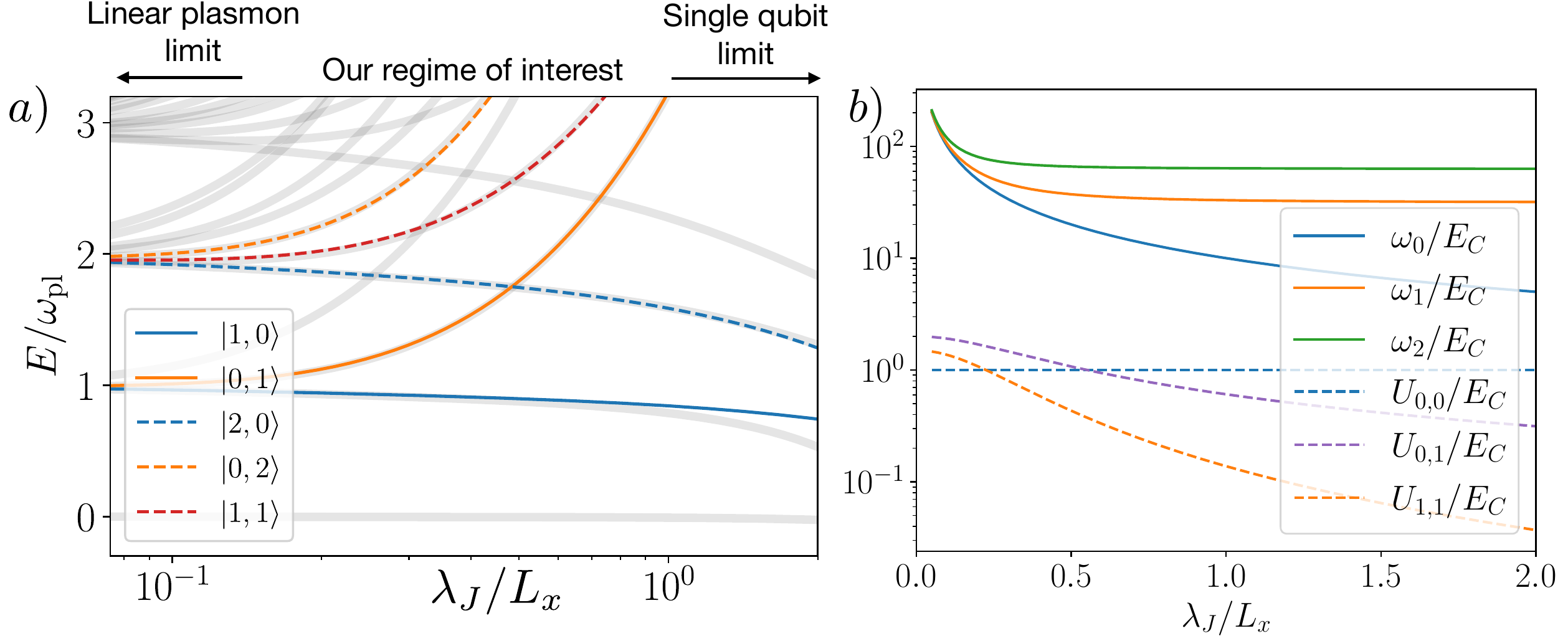} 
\par\end{centering}
\caption{(a) Spectrum of quasi-1D quantum extended Josephson junction as function
of the inverse junction length. Gray lines indicate the numerical
diagonalization of the exact Hamiltonian restricted to three modes.
Colored lines stand for the qubit states of the approximate Hamiltonian
Eq.~\eqref{eq:H2},~\eqref{eq:H4-1}. Zero frequency corresponds
to $\left|0,0\right\rangle $ state. (b) Qubit energies and Hubbard
$U$ of different modes as function of the junction size. We assume
the following parameter ratio $\omega_{\text{pl}}/E_{C}=10$ for $\lambda_{J}/L_{x}$=1.}

\label{Fig3} 
\end{figure}

\twocolumngrid

\section{Extended Josephson junction\label{sec:Long-Josephson-junction}}

We consider a 2D bilayer superconductor, forming an extended Josephson
junction, interacting with the electromagnetic field of a resonator,
as shown in Fig.~\ref{Fig1}~(a). We assume that the size of the
junction is given by $L_{x}\times L_{y}$ with one of the dimensions
($L_{x}$) being comparable to the wavelength of the cavity field.
More importantly, we consider $L_{x}\sim\lambda_{J}$, where $\lambda_{J}$
is the Josephson penetration length that characterizes the stiffness
of phase fluctuations in each superconducting region, and which we explicitly
define below. In conventional qubits $L_{x}\ll\lambda_{J}$, and one
deals with only one Josephson plasmon mode, while in our case, the spatial
dependence of the superconducting phase fluctuations cannot be neglected,
and one can have many Josephson plasmon modes. In this Section, we
derive the complete quantum mechanical description of the multimode
Josephson junction and its interaction with the electromagnetic environment.

As shown in Fig.~\ref{Fig1}~(a), we consider an elongated junction
where $L_{x}\gg L_{y}$. The well-known classical description of the
junction is given by the sine-Gordon Lagrangian \citep{RevModPhys.61.763,T04}:
\begin{equation}
\mathcal{L}_{\text{sG}}=\int\frac{dx}{L_{x}}\left(\frac{1}{16E_{C}}\left(\partial_{t}\theta\right)^{2}-\frac{E_{J}\lambda_{J}^{2}}{2}\left(\partial_{x}\theta\right)^{2}+E_{J}\cos\left(\theta\right)\right),\label{eq:Lsg}
\end{equation}
where $\theta=\theta\left(x,t\right)\equiv\theta_{2}-\theta_{1}+\frac{2e}{c}\int_{1}^{2}{\bf A}d{\bf z}$
denotes the gauge-invariant phase difference between the two superconductors.
$E_{C}$ and $E_{J}$ are the total charging energy and Josephson
energies, respectively, and $\lambda_{J}$ is the Josephson penetration
length $\lambda_{\text{J}}^{2}\equiv\Phi_{0}c/8\pi^{2}(\delta_{z}+2\lambda_{L})j_{c}$,
where $j_{c}$ is the critical current of the junction, $\Phi_{0}$
is the flux quantum, $\delta_{z}$ is the distance between superconducting
islands \citep{T04}. We note that Eq.~\eqref{eq:Lsg} can be derived
from the quantum mechanical description of a bilayer superconductor,
as we demonstrate in App.~\ref{sec:Microscopic-derivation}. The
corresponding Euler-Lagrange equation of motion of the Lagrangian in Eq.~\ref{eq:Lsg}
is known as the sine-Gordon equation:

\begin{equation}
\partial_{x}^{2}\theta-\tilde{c}^{-2}\partial_{t}^{2}\theta=\lambda_{J}^{-2}\sin(\theta),\label{eq:SG_eqn}
\end{equation}
where $\tilde{c}^{2}=8E_{C}E_{J}\lambda_{J}^{2}$. We also supplement
Eq.~\ref{eq:SG_eqn} with the boundary conditions which are $\partial_{x}\theta=0$
at the junction boundary representing the absence of supercurrents
outside the superconducting islands. In this case  the variation of
the phase is small $\delta\theta\ll2\pi$, we can linearize Eqs.~(\ref{eq:Lsg},~\ref{eq:SG_eqn})
which becomes equivalent to a set of harmonic oscillators. Following
this approach, we perform the quantization of the junction below.

\subsubsection{Quantization of phase fluctuations\label{subsec:Quantized-phase-fluctuations}}

We now quantize the fluctuations of phase in the extended Josephson
junction. In the limit when the phase variations are small
$\delta\theta\ll1$, we expand the cosine term in Eq.~\eqref{eq:Lsg}
as $\cos\left(\theta\right)\approx1-\theta^{2}/2+\theta^{4}/24$.
Our goal is now to quantize the phase fluctuations keeping only up
to the quadratic term. We then treat the higher-order terms in Eq.~\eqref{eq:Lsg}
as perturbation. We note that throughout this work we will assume
the junction size is of the order of several Josephson lengths.
In this case, we can safely neglect the spontaneous (thermal) nucleation
of vortices as we discuss in App.~\ref{sec:Nucleation-of-Solitons}.

We expand the phase over a complete set of modes as eigenmodes of
the differential operator $(-\partial_{x}^{2}+\lambda_{J}^{-2})$
with the appropriate boundary condition as $\theta=\sum_{m}\theta_{m}\Xi_{m}\left(x\right)$,
where $\int_{-L_{x}/2}^{L_{x}/2}\Xi_{m}\left(x\right)\Xi_{m'}\left(x\right)dx=L_{x}\delta_{m,m'}$.
By performing the Legendre transformation, we get the classical quadratic
and quartic Hamiltonians:

\begin{align}
H^{(2)} & =\sum_{m}\left\{ 4E_{C}\pi_{m}^{2}+\frac{1}{16E_{C}}\omega_{m}^{2}\theta_{m}^{2}\right\} ,\label{eq:Lsg-1}\\
H^{(4)} & =-\frac{E_{J}}{24}\int\frac{dx}{L_{x}}\left(\sum_{m}\Xi_{m}\left(x\right)\theta_{m}\right)^{4},\label{eq:H4-2}
\end{align}
where $\pi_{m}$ is the momentum and we defined the mode frequency
$\omega_{m}=\omega_{{\rm {pl}}}\sqrt{1+\lambda_{J}^{2}\left(\frac{\pi m}{L}\right)^{2}}$,
and the plasmon frequency $\omega_{{\rm {pl}}}\equiv\sqrt{8E_{J}E_{C}}$.
Eq.~\eqref{eq:Lsg-1} is equivalent to a set of harmonic oscillators
and can perform quantization of harmonic oscillators in Eq.~\eqref{eq:Lsg-1}.
The quantization of Eq.~\eqref{eq:Lsg} can be performed in a standard
way:

\begin{equation}
\hat{H}^{(2)}=\sum_{m}\frac{\omega_{m}}{2}\left(\hat{\pi}_{m}^{2}+\hat{\theta}_{m}^{2}\right)=\sum_{m}\omega_{m}\hat{b}_{m}^{\dagger}\hat{b}_{m},\label{eq:H2}
\end{equation}
where we defined the phase fluctuation and its canonical conjugate
operators are denoted as $\hat{\theta}_{m}=\sqrt{\frac{4E_{C}}{\omega_{m}}}(\hat{b}_{m}^{\dagger}+\hat{b}_{m})$,
$\hat{\pi}_{m}=\sqrt{\frac{\omega_{m}}{16E_{C}}}$$(i\hat{b}_{m}^{\dagger}-i\hat{b}_{m})$
with $\hat{b}_{m}^{\dagger}/\hat{b}_{m}$ being the creation/annihilation
operators of $m-$th mode $[\hat{b}_{m},\hat{b}_{m'}^{\dagger}]=\delta_{m,m'}$.
The zero-point fluctuations of the phase operator are thus given by
$\theta_{\text{ZPF}}=\sqrt{4E_{C}/\omega_{m}}$. For the validity
of the quantization in this section, we need $\theta_{\text{ZPF}}\ll1$,
which is satisfied since we consider $E_{J}\gg E_{C}$.

\subsubsection{Nonlinearity\label{subsec:Nonlinearity}}

We now discuss the quantization of non-linearity of the multiple modes
of Eq.~\eqref{eq:H4-2}. We note that the ambiguity in the operator
ordering can be lifted in the microscopic derivation of the fully
quantum mechanical action of the junction as we present in App.~\ref{sec:Microscopic-derivation}.
Here, for simplicity, we simply extend the derivation of the single-mode
junction \citep{BGG21}, to the multimode case of interest. Neglecting
the excitation non-conserving terms, the nonlinear part of the Hamiltonian
becomes:

\begin{align}
\hat{H}^{(4)} & \approx-\frac{U_{0,0}}{2}\hat{b}_{0}^{\dagger}\hat{b}_{0}^{\dagger}\hat{b}_{0}\hat{b}_{0}-\frac{U_{1,1}}{2}\hat{b}_{1}^{\dagger}\hat{b}_{1}^{\dagger}\hat{b}_{1}\hat{b}_{1}\nonumber \\
 & -U_{0,1}\left(\hat{b}_{1}^{\dagger}\hat{b}_{1}\hat{b}_{0}^{\dagger}\hat{b}_{0}+\frac{1}{4}\hat{b}_{0}^{\dagger2}\hat{b}_{1}^{2}+\frac{1}{4}\hat{b}_{1}^{\dagger2}\hat{b}_{0}^{2}\right).\label{eq:H4-1}
\end{align}
In the limit $L_{x}\sim\lambda_{J}$, the first several terms are
found to be $U_{0,0}=E_{C},$ $U_{0,1}=2E_{C}$ and $U_{1,1}=3E_{C}/2$.
We note that in Eq.~(\ref{eq:H4-1}), we ignored the linear terms
that renormalize the plasmon frequencies. In the case of a finite
ratio $L_{x}/\lambda_{J}$, the nonlinearity coefficients depend on
the junction size as shown in Fig.~\ref{Fig3}(a). The three terms
in the equation above respectively represent the self-, cross-Kerr,
and parametric interaction terms. As shown in Fig.~\ref{Fig3}(b),
the strengths of these various different terms strongly depend on
the size of the junction. For larger junction sizes, the spacing between
qubit frequencies is decreasing, and the interaction strengths ($U$'s)
are decreasing. We note that the charge sensitivity \citep{KTG07}
is inversely proportional to the phase uncertainty, and for our choice
of parameters (also known as the transmon regime $E_{J}\gg E_{C}$)
it is negligible.

In addition to the Kerr and cross-Kerr terms in Eq.~\eqref{eq:H4-1},
the terms involving three and four different modes are allowed by
the symmetry of the junction $\hat{H}^{\prime(4)}\approx-U_{0,1,2}\hat{X}_{2}\hat{X}_{1}^{2}\hat{X}_{0}$,
$\hat{H}^{\prime\prime(4)}\approx-U_{0,1,2,3}\hat{X}_{3}\hat{X}_{2}\hat{X}_{1}\hat{X}_{0}$
with $\hat{X}_{i}\equiv b_{n}+b_{n}^{\dagger}$. In the limit $L_{x}/\lambda_{J}\gg1$,
we find $\chi_{0,1,2}\approx E_{C}/\sqrt{2}$ and $\chi_{0,1,2,3}\approx E_{C}\sqrt{2}$.
We note that the plasmon frequencies $\omega_{m}$ are generally incommensurate,
and therefore $H'^{(4)}$ and $H''^{(4)}$ terms are excitation non-preserving.
However, one can make these terms resonant by driving, where the energy
mismatch is compensated by an external driving as we discuss in the
Sec.~\ref{sec:Quantum-simulation-of}.

\section{Coupling to the electromagnetic field\label{subsec:Coupling-to-the}}

We proceed to examine the interaction between the extended JJ
modes, derived in the previous Section, and the electromagnetic field
of a microwave resonator, as depicted in Fig.~\ref{Fig1}~(a). Our
focus is on the interaction with the $E^{(z)}$ field component, which induces a voltage across the junction. In the classical
picture, the external voltage bias can be taken into account on the
level of Lagrangian formulation Eq.~\eqref{eq:Lsg} by the following
substitution $\partial_{t}\theta\rightarrow\partial_{t}\theta+2e\int_{1}^{2}dzE^{(z)}$.
By performing the Legendre transformation of the resulting Lagrangian
as discussed in App.~\ref{qubit-light coupling}, we get the classical
coupling Hamiltonian $H_{\text{int}}=2e\delta_{z}\sum_{n}\pi_{m}E_{m}^{(z)}$,
where the overlap of the electric field and the $m$-th plasmon mode
profile is defined as $E_{m}^{(z)}\equiv\int\frac{dx}{L_{x}}E^{(z)}\left(x\right)\Xi_{m}\left(x\right)$.
In the following, we assume this coupling is weak such that we can separately quantize the electromagnetic field
and the superconductor.

Initially, we examine the quantization of transverse
electromagnetic (TEM) modes in the electromagnetic field of a coplanar resonator
shown in Fig.~\ref{Fig1}~(a). We assume the resonator length along
$x$ direction as $L^{\text{res}}$. We do not discuss
the quantization procedure as it can be found in e.g. \cite{BGG21}.
The quantized operator of the electromagnetic field takes the standard
form \citep{KK94, GAF10}:

\begin{equation}
\hat{E}^{(z)}(x)=\sum_{n}E_{n}^{0}(i\hat{a}_{n}{\cal E}_{n}\left(x\right)+\text{H.c.}),\label{eq:Ez}
\end{equation}
where ${\cal E}_{n}$$\left(x\right)$, $\omega_{n}^{\text{res}}$,
respectively denote the electromagnetic mode profile and the frequency
of $i$-th resonator mode. $E_{n}^{0}$ is the zero-point electric field of $n$-th mode of the resonator. In this
section, we restrict our consideration to a homogeneous field distribution
along $y$ and $z$ directions in Eq.~\eqref{eq:Ez} for simplicity.
The total Hamiltonian of the EM field is given by $H_{\text{EM}}=\sum_{n\geq0}\omega_{n}^{\text{res}}a_{n}^{\dagger}a_{n}$
with $\omega_{n}^{\text{res}}=\pi(n+1)c/L^{\text{res}}$. In Eq.~\eqref{eq:Ez},
we only considered the electromagnetic field with a
polarization component along $\hat{z}$ direction that induces a voltage across
the junction. Here we assume the zero voltage boundary conditions which can be
achieved by grounding the end points of the resonator \citep{BKH13}.

The quantized coupling to the Josephson junction can be obtained from
the classical Hamiltonian $H_{\text{int}}$ by simply replacing fields
with the corresponding operators. Ignoring the excitation-non-preserving
terms, we get:

\begin{equation}
\hat{H}_{\text{int}}=\sum_{m,n}g_{m,n}\hat{a}_{n}\hat{b}_{m}^{\dagger}+\text{H.c.},\label{eq:Hcoupl}
\end{equation}
where the qubit-photon couplings are $g_{m,n}=2e\delta_{z}E_{n}^{0}\sqrt{\frac{\omega_{m}}{16E_{C}}}\int\frac{dx}{L_{x}}{\cal E}_{n}\left(x\right)\Xi_{m}\left(x\right)$
and $\delta_{z}$ denotes the distance between the superconducting
layers.  We note that our description of light-matter coupling is a simplified one, and in realistic scenarios, the effective thickness $\delta_z$ 
 is determined by a combination of many factors, including the geometry of capacitors, and it is not equal to the distance between superconducting islands.  The most reliable approach for determination of the coupling coefficients $g_{m,n}$ is based on the finite element methods \cite{KTG07}.

These couplings can be evaluated in a general case but it is useful
to consider a configuration where some of the couplings vanish for symmetry reasons. In particular, if the junction is placed
in the center of the resonator, some of the coupling coefficients
vanish due to the selection rules imposed by the inversion symmetry
with respect to $x$-axis. Specifically, we have $g_{m,n}\neq0$,
if $m$, $n$ are both odd or even. The effective circuit description
of such a junction-resonator configuration is shown in Fig.~\ref{Fig1}~(b).
The multimode structure of an extended junction enables the encoding
of multiple qubits within a single junction. Mode-selective coupling
to the electromagnetic field serves as a versatile tool for manipulating
the quantum state, as depicted in Fig.~\eqref{Fig1}~(c), where
we schematically illustrate all the possible couplings between the
Fock states of the junction and the resonator allowed for by Eq.~\eqref{eq:Hcoupl}.
On a semiclassical level, our coupling can be used to characterize
the nonlinearity of the junction itself as discussed in App.~\ref{subsec:Experimental-characterization-pr}.

\section{Quantum simulation with extended Josephson junctions\label{sec:Quantum-simulation-of}}

In the following section, we aim to illustrate the wide-ranging capabilities
of our setup. We show that our system functions as a complete toolbox
for simulating many-body physics and lattice gauge theories. In particular,
we examine a system of $N$ extended Josephson junctions interacting with
the electromagnetic field within a resonator, as depicted in Fig.~\ref{Fig4}~(a).
We further assume that these junctions are subjected to external driving,
and as we will demonstrate later, this can result in excitation tunneling
in a synthetic dimension that is parameterized by $m$. We focus exclusively
on the interaction of each junction with the resonator, neglecting
potential direct capacitive (dipole-dipole) couplings between the
junctions. This assumption is valid as long as the distance between
any pair of junctions is greater than the distance between JJs and
the waveguides. Expanding to a multi-junction setup and incorporating
their connection to the waveguide can be achieved, by adhering to
the derivation outlined in Sec.~\eqref{subsec:Coupling-to-the}.
By denoting the creation/annihilation operators of the $i$-th junction
by $b_{m}^{(i)\dagger}/b_{m}^{(i)}$, we express the full Hamiltonian
$H=H_{\text{q}}+H_{\text{nl}}+H_{\text{tun}}+H_{\text{tun}}^{\prime}$:

\begin{align}
H_{\text{q}} & =\sum_{m}\omega_{m}b_{m}^{(i)\dagger}b_{m}^{(i)},\label{eq:Hqubit}\\
H_{\text{nl}} & =-\sum_{m,n,i}U_{m,n}\hat{b}_{m}^{(i)\dagger}\hat{b}_{n}^{(i)\dagger}\hat{b}_{n}^{(i)}\hat{b}_{m}^{(i)},\label{eq:Hnl}\\
H_{\text{tun}} & =-\sum_{i,j,m}t_{i,j}^{(m)}b_{m}^{(i)\dagger}b_{m}^{(j)}+\text{H.c.},\label{eq:Htun}\\
H_{\text{tun}}^{\prime} & =-\sum_{i,m,m'}t_{i}^{\prime(m,m')} b_{m}^{(i)\dagger}b_{m'}^{(i)}+\text{H.c.},\label{eq:Htun1}
\end{align}
whereas before we ignore the excitation-non-preserving terms. The
first two terms $H_{{\bf q}}$ and $H_{\text{nl}}$ stand for the
Hamiltonians of each junction while $H_{\text{tun}}$ and $H_{\text{tun}}^{\prime}$
correspond to the different excitation tunneling processes that will
be discussed below.

\onecolumngrid 
\begin{center}
\begin{figure}
\begin{centering}
\includegraphics[scale=0.5]{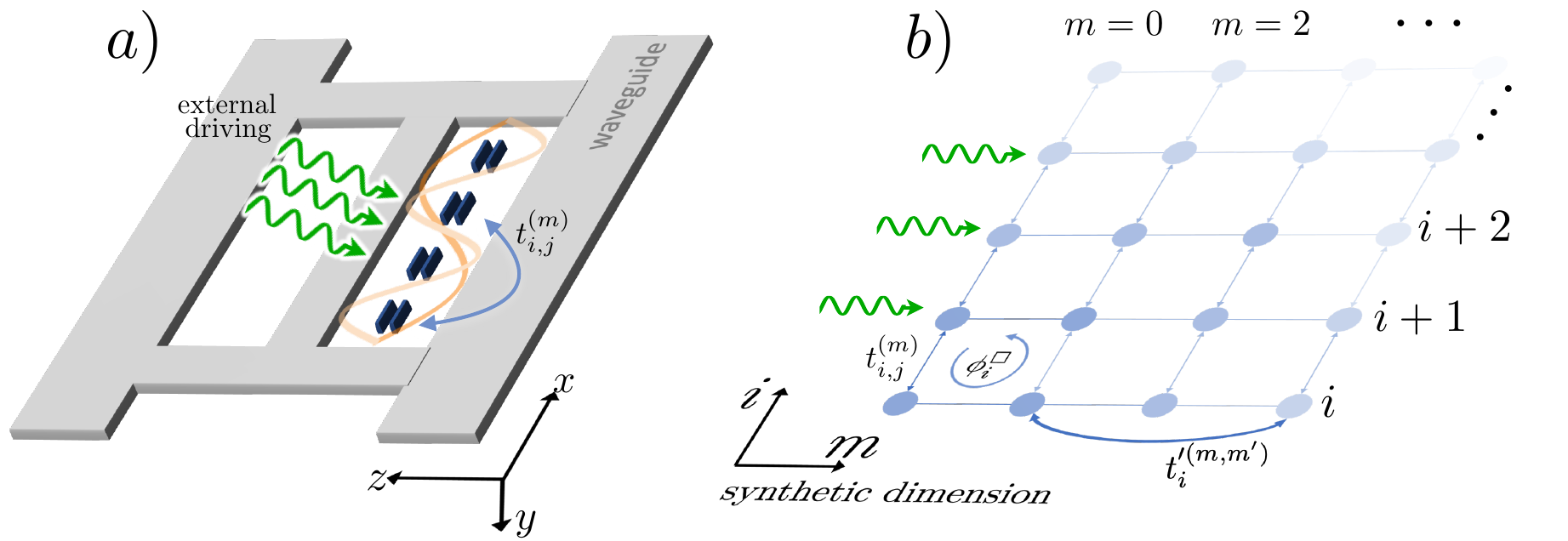} 
\par\end{centering}
\caption{Implementation of multi-component Bose-Hubbard model with an array
of extended Josephson junctions. (a)  Schematic representation of an array of 4
extended JJ is placed in a resonator.  $n=2$ and $n=4$ resonator modes are shown in orange and pale orange respectively. Each junction is placed in an anti-node of $n=4$ mode. (b) Equivalent
Bose-Hubbard model in synthetic dimension. For schematic simplicity the junctions are shown with layers stretched along $z$ direction, i.e. along the dominant electric field of the resonator. The actual experimentally motivated geometry is shown in Fig.~\ref{Fig_realistic}.}

\label{Fig4} 
\end{figure}
\par\end{center}

\twocolumngrid

We begin with a discussion of the Hamiltonian equation denoted as
Eq.~\eqref{eq:Htun1}, which corresponds to the tunneling of an excitation
between different junctions, mediated by the waveguide. The microscopic
representation of this process is encapsulated by the Hamiltonian
$H_{\text{int}}=\sum_{m,n,i}g_{m,n}^{(i)}a_{n}b_{m}^{(i)\dagger}+\text{H.c.}$
which is an elaboration of Eq.~\eqref{eq:Hcoupl}, adapted to accommodate
scenarios involving multiple junctions. Adiabatically eliminating
the resonator modes, we get Eq.~\eqref{eq:Htun1} with the detuning
denoted as $\delta_{m,n}=\omega_{m}-\omega_{n}^{\text{res}}$ and
$t_{i,j}^{(m)}=\sum_{n}g_{m,n}^{(j)}g_{m,n}^{(i)}/\delta_{m,n}$,
where the summation is taken over all the resonator modes parametrized
by $n$.  We note that $t_{i,j}^{(m)}$ is proportional to the Green's
function describing the photon propagation between two junctions $i$
and $j$, where its specific form depends on the waveguide configuration. For example, in the configuration shown in Fig.~\eqref{Fig4}~(a),
the tunneling coefficient $t_{i,j}^{(m)}$ can be expected to be of
long- (infinite) range character. We note that the tunneling can become
short-range in the case of each junction coupled to its own resonator in a resonator array
as in \citep{HTK12, SK13,SLZ19,MKF18}.  To maximize tunneling and consequently the achievable quantum simulation times, it is beneficial to minimize the detuning between plasmon and cavity modes. One approach for achieving this is through the adjustment of the junction geometry. Specifically, selecting the appropriate junction length can adjust the spacing between plasmon modes, as follows from Eq.~\eqref{eq:Wm}. Alternatively, the plasmon frequencies can be adjusted via external driving such as to induce AC Stark frequency shifts.

The last term, $H_{\text{tun}}^{\prime}$, characterizes the resonant excitation exchange among different plasmon modes inside one junction, which, as we show below, can be induced by an external driving. As a specific example, we now outline a protocol for implementing the tunneling between $m=0$ and $m=2$ junction modes, keeping in mind that a similar approach could be used to induce tunneling between other odd pairs of modes. In particular, we consider a weak
drive for the $m=1$ mode described by the RWA Hamiltonian $H_{\text{dr}}(t)=\alpha_{1}\sum_{i}(\hat{b}_{1}^{(i)}e^{i\omega_{1}^{\text{dr}}t+i\phi_{i}}+\hat{b}_{1}^{(i)\dagger}e^{-i\omega_{1}^{\text{dr}}t-i\phi_{i}}$)
with $\omega_{1}^{\text{dr}}=(\omega_{2}-\omega_{0})/2$ and $\phi_{i}$
denotes the driving phase of $i$-th junction and $\alpha_{1}$ is
the effective Rabi frequency. The selective excitation of a single mode can be realized through the momentum or frequency specificity of the coupling. This yields the mean-field quasi-steady-state
coherence of the driven mode $\bar{b}_{1}\approx-\alpha_{1}e^{i\phi_{i}}/(\omega_{1}^{\text{dr}}-\omega_{1})$.
With this, the tunneling between plasmon modes is induced by the non-linear
excitation-non-conserving Hamiltonian $\hat{H}^{\prime(4)}$ (see
Sec.~\ref{subsec:Nonlinearity}) that is now driven. We get an effective
resonant tunneling $0\leftrightarrow2$ that is given by Eq.~\eqref{eq:Htun1}
with $t_{i}^{(0,2)}=U_{0,1,2}|\bar{b}_{1}|^{2}e^{-2\phi_{i}}$. This
shows that the phase of complex tunneling between same-parity sites
is set by drive. Extending $H_{\text{tunn}}^{\prime}$ to all the
sites and combining it with $H_{\text{tunn}}$, we find a rich set
of tunneling terms in 2D shown in Fig.~\ref{Fig4}~(b), where one
dimension is spatial JJ sites and the other dimension is different
plasmon modes, in form of a synthetic dimension. In particular, one
can set the drive phases to achieve a non-zero total tunneling phase
over a plaquette: $\phi_{i}^{\boxempty}=2\left(\phi_{i}-\phi_{i+1}\right)$ to
obtain an artificial gauge field. We summarize by mentioning 
that our final Hamiltonian is given by Eqs.~(\ref{eq:Hqubit}-\ref{eq:Htun1})
and it represents the multi-component Bose-Hubbard model in an artificial
external magnetic field.

The multicomponent Bose-Hubbard model enables the implementation of
quantum simulation protocols for intricate many-body phenomena, which
we discuss it here. It is worth noting that comparable models
have also emerged in the study of conventional Josephson junction
arrays as referenced in \citep{VFE92,MSS01}, as well as in the examination
of arrays of coupled cavities inhabited by multi-level atoms \citep{HBP08,KKC08}.
This model is recognized for exhibiting phenomena such as spin components
\textquotedbl{}demixing\textquotedbl{} \citep{CH03} and spin-charge
separation \citep{RFZ03}. When subjected to strong interactions,
it is anticipated that the Hubbard models will transition to the Mott-insulator
state \citep{CZP12}. In this state, the low-energy dynamics are characterized
by the effective spin Heisenberg models represented in different bosonic
components \citep{HBP08}. Presenting this system to a synthetic gauge
field, as in our situation, can reveal a spectrum of fascinating phenomena,
including a Mott insulator state with chiral currents \citep{DMM12}
and aspects of quantum Hall physics \citep{hafezi2007fractional,TG15}.
Furthermore, the glassy phases predicted in a single-component polaritonic
Hubbard model are explored in \citep{RF07}. We finish this section
by stating that the effective Hubbard model, as presented by Eqs.~(\ref{eq:Hqubit}-\ref{eq:Htun1}),
possesses all the necessary elements to perform quantum simulations
of lattice gauge theories \citep{abb22}, as recently suggested in
\citep{OMY22, MZH20}.

\section{Qubit operation \label{sec:qubit}}

We now discuss the potential application of our setup to implement
single and two-qubit gates for quantum computation. We assume the
individual addressing of each junction mode by using the parity selection
rules as discussed above. The single-qubit rotations can be performed
trivially by applying the external microwave tone of the corresponding frequency.
Therefore below, we focus only on two-qubit gates. Theoretically,
the phase gate can be implemented for any pair of qubits residing
in one or separate junctions. Below we focus on the former case as
we expect it to have much higher fidelity.

\paragraph*{Two-qubit gate ---}

\begin{figure}
\begin{centering}
\includegraphics[scale=0.4]{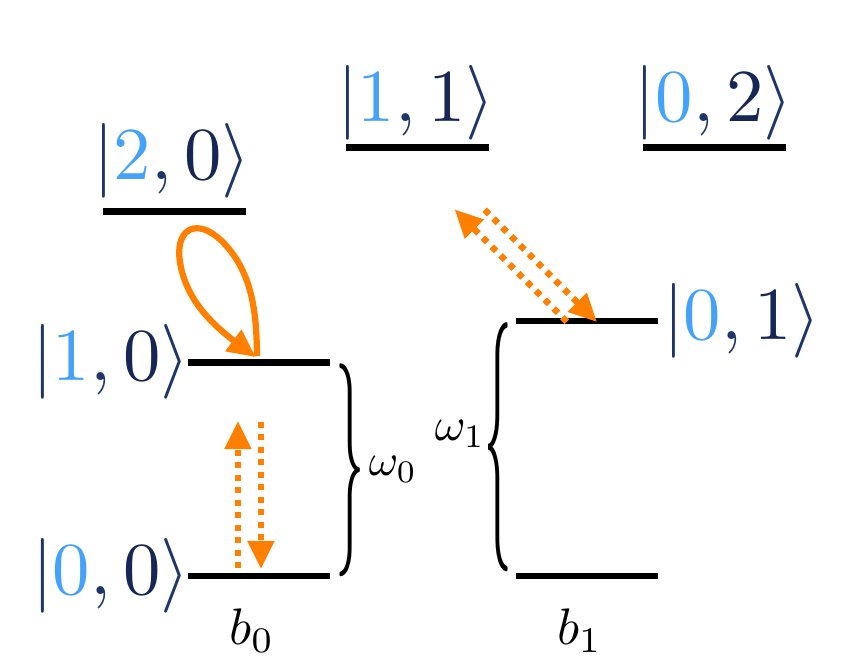} 
\par\end{centering}
\caption{Implementation of an intra-junction two-qubit phase gate. The qubit
$b_{0}$ is driven resonantly with the transition $\left|1,0\right\rangle \rightarrow\left|2,0\right\rangle $
for half a period of a corresponding Rabi oscillation picking up a
$\pi$ phase. Other transitions are assumed to be off-resonant due
to the Kerr and cross-Kerr interaction terms between different junction
modes. }

\label{Fig_gate} 
\end{figure}

We now exemplify the implementation of an intra-junction two-qubit
gate ignoring the coupling to the resonator. We operate here
with the logical states $|0\rangle,|1\rangle$ of the two lowest modes
of the junction $b_{0}$ and $b_{1}$, as shown in Fig.~\ref{Fig1}~(c).
We define the two-qubit phase gate as the unitary transformation:
$|i,j\rangle\rightarrow\exp{(i\pi\delta_{i,1}\delta_{j,0})}|i,j\rangle$,
where $\delta$ is the Kronecker delta symbol. Such a unitary can
be implemented by assuming a field driving that is resonant only
with the transition $\left|1,0\right\rangle \rightarrow\left|2,0\right\rangle $.
By driving this transition for half of the period of a full Rabi
oscillation, the junction state picks up a $\pi$ phase $\left|1,0\right\rangle \rightarrow e^{i\pi}\left|1,0\right\rangle $.
In a realistic situation the two other symmetry-allowed transitions
$\left|1,0\right\rangle \rightarrow\left|0,0\right\rangle $ and $\left|0,1\right\rangle \rightarrow\left|1,1\right\rangle $
will be driven as well as shown in Fig~\ref{Fig_gate}. The detunings
of these two transitions are respectively given by the nonlinearity
terms $U_{0,0}$ and $U_{1,1}\approx2U_{0,0}$ in the limit of large
junction. Therefore the estimate of the relevant degree of qubit non-linearity can be
given in terms of the ratio of the anharmonicity to the qubit decoherence
rate $U_{0,0}/\gamma_{\text{d}}$. The estimate of this ratio is provided
in the section below.

\section{Experimental considerations}\label{sec:exp_cons}

We now discuss the experimental parameters of the extended junction
described above. Our estimations are based on the physical parameters
in Ref.~\citep{MHC21}. Although a square geometry is employed in
that reference, we assume that similar parameters apply to our rectangular
geometry. In this section, we will use SI units for experimental purposes,
deviating from the Gaussian units employed throughout the rest of
the text.

Using the critical current value $j_{c}\approx10\text{nA/\ensuremath{\mu\text{m}^{2}}}$,
where $\epsilon_{0}$ is the vacuum permittivity, and assuming the
oxide thickness $\delta_{z}\approx2\text{nm}$ and the London penetration
length of aluminum $\lambda_{L}=16\text{nm}$, we calculate the Josephson
length to be $\lambda_{J}\approx\sqrt{\Phi_{0}/2\pi\mu_{0}(\delta_{z}+2\lambda_{L})j_{c}}\approx880\mu\text{m}$,
where $\Phi_{0}$ denotes the flux quantum \citep{PFC13}.

For the junction area, we assume the following geometric parameters:
$S\approx10\text{nm}\times\lambda_{J}$. Using the above critical
current density, we obtain the Josephson energy as $E_{J}=Sj_{c}/2e\approx2\pi\times40\text{GHz}$.
The charging energy can be determined as $E_{C}\approx\omega_{\text{pl}}^{2}/8E_{J}\approx2\pi\times0.06\text{GHz}$,
assuming that the fundamental plasmon frequency is not influenced
by the geometry and has a value of $\omega_{\text{pl}}\approx2\pi\times5\text{GHz}$.
The cavity qubit coupling in our toy model can be estimated using the experimentally achievable value of the zero-point electric field $E_{n=0}^{0}=0.2\text{V/m}$ \citep{WSB04}. With this we find the single-photon Rabi frequency $g_{0,0}\approx3$ MHz (see Appendix \ref{qubit-light coupling}).

\begin{figure}[h]
\begin{centering}
\includegraphics[scale=0.4]{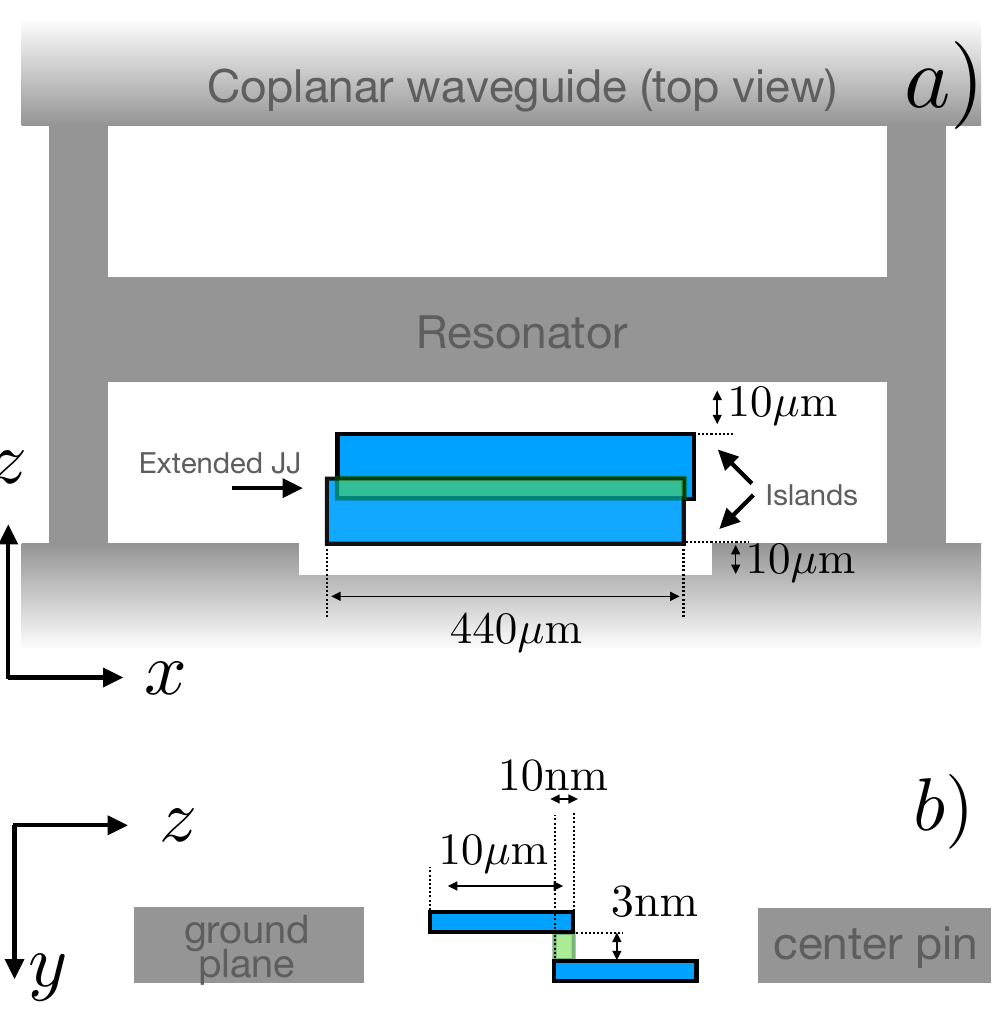} 
\par\end{centering}
\caption{Schematic (not to scale) representation of the long Josephson junction configuration used for the finite-element numerical simulation. a) top view, b) side view. Regions highlighted in green denote the junction area.}

\label{Fig_realistic} 
\end{figure}

We also compare our predictions with the results of finite-element numerical simulation using the commercially available software (Ansys Maxwell) for a more realistic geometry. More precisely, we consider two superconducting islands of dimensions $10\times440 \mu \text{m}^2$ oriented in the x-z plane one on top of the other with the overlap junction area of $10 \text{nm}\times440 \mu \text{m}$ as shown in Fig.~\ref{Fig_realistic} a), b). With this we find the following junction parameters $E_C=2\pi\times 84$ MHz, $E_J\approx2 \pi \times 22$ GHz, $\omega_{\text{pl}}\approx 2 \pi \times3.8$ GHz. We also estimate the qubit - cavity coupling strength to be $g_{0,0}\approx2\pi\times 24$ MHz. The enhanced coupling strength in this geometry stems from the fact that the superconducting islands are larger than the JJ itself and contribute to effective thickness $\delta_z$ \citep{KTG07}

As mentioned earlier, larger Josephson junctions (JJs) have smaller
nonlinearities and one needs to ensure that the nonlinearity strength
(approximately $E_{C}$) remains larger than the dissipation rates.
Assuming an energy dissipation time scale of $T_{1}\approx30\mu s$,
we estimate such a ratio to be favorable, with $2\pi E_{C}T_{1}~\sim10^{4}$.

We note that for larger junction sizes, where $L\sim\lambda_{J}$,
an additional imperfection source arises from Josephson vortex nucleation.
The corresponding vortex density in equilibrium is given by \citep{W80,KM89}:

\[
\left\langle n\right\rangle =\left(\frac{2}{\pi}\right)^{1/2}\lambda_{J}^{-1}\sqrt{\frac{2E_{J}}{k_{{\rm B}}T}}e^{-2E_{J}/k_{{\rm B}}T}.
\]
For the parameters considered, $\left\langle n\right\rangle L_{x}\ll1$,
which indicates that nucleation can be neglected.

\section{Conclusions \& outlook}

In this work, we developed a theoretical framework for describing
the light-matter interaction in extended Josephson junctions. We demonstrated
that each such junction could host multiple plasmon modes, each encoding
a qubit. It is possible to address each qubit mode individually due
to the frequency-momentum-selective coupling, which arises from the
different profiles of their electronic wavefunctions. We also consider
the system of several extended junctions interacting through a single
resonator and demonstrate that such a system could be used to simulate
an interacting 2D Bose-Hubbard model with one of its dimensions being
synthetic. Besides, we also show that the system can be used in order
to perform quantum computation. In particular, we proposed the implementation
of single- and two-qubit gates inside a single extended Josephson
junction. Our work allows us to address several interesting problems
in the future, including the possibility of inducing photon-photon
interactions in long junctions and generating non-classical states
of light. Another interesting direction would be to study the effects
of non-trivial junction geometry (topology) on light-matter interactions.

\section*{Acknowledgments}

The authors thank Z. Minev and M. Devoret for the fruitful discussions. We also thank Maya Amouzegar for help with executing the numerical simulations presented in this work. This material is based upon work supported by the U.S. Department of Energy, Office of Science, National Quantum Information Science Research Centers, and Quantum Systems Accelerator. Additional support is acknowledged from AFOSR MURI FA9550-19-1-0399, FA9550-22-1-0339, ARO W911NF2010232 and NSF QLCI OMA-2120757.

 \bibliographystyle{apsrev4-1}
\bibliography{biblio}

\begin{widetext}

\appendix

\section{Microscopic derivation\label{sec:Microscopic-derivation}}

In this section we derive the effective action of Josephson junction
interacting with the resonator EM field. The setup we have in mind
is shown in Fig.~\ref{Fig1}. In this note to some extent the formalism
follows \citep{AS10,SFB20}. We consider the imaginary-time action
of two quasi-two-dimensional superconductors $S_{\text{full}}=S_{0}+S_{J}+S_{\text{EM}}$:

\begin{align}
S_{0} & =\sum_{i=1,2}\int d^{2}{\bf r}L_{z}\int_{0}^{\beta}d\tau\left\{ \frac{\nu}{2}\left(\partial_{\tau}\theta_{{\bf r}}^{(i)}-e\phi_{{\bf r}}^{\left(i\right)}\right)^{2}+\frac{n_{s}}{2m}\left(\nabla\theta_{{\bf r}}^{\left(i\right)}-\frac{e}{c}{\bf A}_{{\bf r}}^{\left(i\right)}\right)^{2}\right\} ,\label{eq:S0}\\
S_{\text{J}} & =-\frac{E_{\text{J}}}{S}\int d^{2}{\bf r}\int_{0}^{\beta}d\tau\cos2\left(\theta_{{\bf r}}^{(1)}-\theta_{{\bf r}}^{\left(2\right)}-\frac{e}{c}\int_{1}^{2}d\vec{z}\cdot{\bf A}_{{\bf r},z}\right),\label{eq:SJ-1}
\end{align}
where $\phi_{{\bf r}}^{(i)},{\bf A}_{{\bf r}}^{(i)}$, denotes the
values of the scalar and vector potentials at $i$-th layers respectively,
$m,e$ are the electron mass and charge respectively, $n_{s}$ denotes
the superfluid density. $\theta^{(i)}$ is the phase of $i$-th superconductor.
The boundary conditions imposed by resonator can be straightforwardly
included into Eqs.~\ref{eq:S0}. In the following we denote the in-plane
vectors as ${\bf r}$. The size of each superconductor is assumed
to be $L_{x}\times L_{y}\times L_{z}$ such that $L_{z}\ll L_{x,y}$.
$E_{\text{J}}$ is the Josephson coupling energy.

We follow the procedure outlined below. We first consider the junction
alone and derive its effective action by integrating-out the static
components of EM field responsible for the inductive and charging
interaction between superconductors. We then add the interaction with
the electromagnetic resonator modes perturbatively in a gauge-invariant
way with the only assumption that resonator does not affect the Lorentz
and Coulomb force between superconductors.

\subsection{Josephson junction action}

We now define the symmetric and the anti-symmetric phase variable
as follows $\theta_{{\bf r}}^{\left(\pm\right)}=2\left(\theta_{{\bf r}}^{(1)}\pm\theta_{{\bf r}}^{(2)}\right)$
and we integrate-out the static parts of vector and scalar potentials
(see below) in Eqs.~(\ref{eq:S0}-\ref{eq:SJ-1}). We find the following
action:

\begin{align}
S_{0} & \approx\int\frac{d^{2}{\bf r}}{S}d\tau\left\{ \frac{1}{16E_{C}}\left(\partial_{\tau}\theta_{{\bf r}}^{(-)}\right)^{2}+\frac{E_{\text{J}}\lambda_{\text{J}}^{2}}{2}\left(\vec{\nabla}\theta_{{\bf r}}^{\left(-\right)}\right)^{2}\right\} \label{eq:S0-1}\\
S_{\text{J}} & \approx-\frac{E_{\text{J}}}{S}\int d^{2}{\bf r}d\tau\cos\left(\theta_{{\bf r}}^{(-)}\right)\label{eq:SJ}
\end{align}
where $E_{C}$ is the total charging energy and $\lambda_{\text{J}}^{2}\equiv\Phi_{0}c/8\pi^{2}\delta_{z}j_{c}$
is the Josephson penetration length and $S$ is the surface are of
the junction. $\Phi_{0}$ denotes the flux quantum and $j_{c}$ is
the critical current through the junction which we assume to be constant.
The $+$ component of the field becomes gapped and is neglected
in the action above. We note that the Josephson term \ref{eq:SJ} is
only valid at lengthscales larger than the coherence length $k^{-1}\gg v_{F}/\Delta$
where $\Delta$ is the gap of the superconductor and $v_{F}$ is the
Fermi velocity.

\subsection{Quantized phase fluctuations}

In this section we derive the Hamiltonian of the effective Josephson
qubit. We first expand the action Eqs.~\ref{eq:S0-1}-\ref{eq:SJ}
up to quartic term and get:

\begin{align}
S^{\left(2\right)} & \approx\frac{1}{16E_{C}S}\int d^{2}{\bf r}d\tau\theta_{{\bf r}}^{\left(-\right)}\left\{ -\partial_{\tau}^{2}+\omega_{\text{pl}}^{2}\left\{ 1-\lambda_{\text{J}}^{2}\vec{\nabla}^{2}\right\} \right\} \theta_{{\bf r}}^{\left(-\right)}\label{eq:S2}\\
S^{\left(4\right)} & \approx-\frac{1}{24}\frac{E_{\text{J}}}{S}\int d^{2}{\bf r}d\tau\theta_{{\bf r}}^{(-)4}\label{eq:S4}
\end{align}
where the Josephson plasmon frequency is $\omega_{\text{pl}}=\sqrt{8E_{\text{C}}E_{\text{J}}}$.
If we neglect the fact that $\theta$ is the angle variable and treat
it as a bosonic field then the linear part of the action Eq.~\ref{eq:S2}
represents the set of harmonic oscillators. We now expand the phase
field over eigensystem of the $\left\{ 1-\lambda_{\text{J}}^{2}\vec{\nabla}^{2}\right\} $
operator with the boundary conditions $\nabla\theta^{\left(-\right)}=0$:

\begin{equation}
\theta_{{\bf r}}^{\left(-\right)}\left(\tau\right)\equiv\sum_{m}\sqrt{\frac{4E_{\text{C}}}{\omega_{m}}}\Xi_{m}\left({\bf r}\right)\theta_{m}\left(\tau\right),\label{eq:theta}
\end{equation}
where $\left\{ 1-\lambda_{\text{J}}^{2}\vec{\nabla}^{2}\right\} \Xi_{m}\left({\bf r}\right)\equiv\epsilon_{m}^{2}\Xi_{m}\left({\bf r}\right)$
and we assumed the normalization condition $\int d^{2}{\bf r}\Xi_{m}\left({\bf r}\right)\Xi_{m'}\left({\bf r}\right)=S\delta_{m,m'}$, where $S$ is the junction surface.
The prefactor in Eq.~\eqref{eq:theta} guarantees the proper normalization
of the phase fluctuations. From now on we focus on a ``long'' rectangular
Josephson junction, i.e. we consider $L_{x}\gg L_{y}$. Assuming $x\in[\frac{L_{x}}{2},\frac{L_{x}}{2}]$
we can restrict the modes to $\Xi_{m}(x,y)\approx\Xi_{m}\left(x\right)=\sqrt{2}\sin\frac{\pi x}{L_{x}}m$
for $m$ being odd, $\Xi_{m}\left(x\right)=\sqrt{2}\cos\frac{\pi x}{L_{x}}m$
for m being even and $\Xi_{m=0}\left(x\right)=1$. The oscillator
energies are given by 
\[
\omega_{m}=\omega_{\text{pl}}\sqrt{\lambda_{\text{J}}^{2}\left(\frac{\pi}{L_{x}}m\right)^{2}+1}.
\]
Substituting Eq.~\eqref{eq:theta} to Eq.~\eqref{eq:S2} we can
infer the effective linear Hamiltonian (we just replace the fluctuating
field with bosonic variables as $\theta_{m}\rightarrow b_{m}+b_{m}^{\dagger}$,
where $\left[a_{m},a_{m'}^{\dagger}\right]=\delta_{m,m'}$ ):

\[
H^{(2)}=\sum_{m}\epsilon_{m}b_{m}^{\dagger}b_{m}
\]
The quartic term depends on the junction parameters. Below we derive
the expression for the junction assuming $L_{x}\gg L_{y}$ for simplicity.

\subsection{Extended Josephson junction}

We now derive the nonlinear term for the junction assuming $L_{x}\gg L_{y}$.
We note that the quadratic part of the Hamiltonian corresponding to
Eq.~\eqref{eq:S4} has divergent contributions. In particular, the
correction to the 0-th plasmon mode has the following form $E_{C}\left(\sum_{n\geq0}1/\sqrt{1+\lambda_{J}^{2}\left(\frac{\pi n}{L}\right)^{2}}\right)$,
where the sum taken is over all modes. The contribution to the rest
modes is found to be $1/\sqrt{1+\lambda_{J}^{2}\left(\frac{\pi n}{L}\right)^{2}}E_{C}\left(\frac{1}{2}+\sum_{m\geq0}1/\sqrt{1+\lambda_{J}^{2}\left(\frac{\pi m}{L}\right)^{2}}\right)$.
The mode cut-off has to be imposed to get a convergent expression.
Such cut-off can be obtained on the physical grounds by e.g. demanding
the momentum of the contributing mode to be smaller than the coherence
length of the superconductor. Here we simply absorb the formally divergent
sum into the plasmon frequency definition. In this case, the result
is cut-off independent. Below we restrict our consideration to just
the two lowest plasmon modes.

Restricting to the lowest two modes we find the expression for the
quartic term corresponding to the nonlinear part of the action Eq.~\eqref{eq:S4}:

\begin{align}
H^{(4)} & =\frac{E_{C}}{12}\int_{0}^{L_{x}}\frac{dx}{L_{x}}\left\{ \sqrt{\frac{\omega_{\text{pl}}}{\epsilon_{0}}}X_{0}+\sqrt{\frac{2\omega_{\text{pl}}}{\epsilon_{1}}}\cos\left(\frac{\pi}{L_{x}}x\right)X_{1}\right\} ^{4}\nonumber \\
 & =\frac{E_{C}}{12}\left\{ X_{0}^{4}+6\frac{\omega_{\text{pl}}}{\epsilon_{1}}X_{0}^{2}X_{1}^{2}+\left(\frac{\omega_{\text{pl}}}{\epsilon_{1}}\right)^{2}\frac{3X_{1}^{4}}{2}\right\} \label{eq:H4}
\end{align}
where we neglected quadratic terms and $X_{i}\equiv b_{i}+b_{i}^{\dagger}$.
The Hamiltonian Eq.~\eqref{eq:H4} can be simplified even further
assuming that the non-linearity is weak compared to the plasma frequency
$\omega_{\text{pl}}\ll E_{C}$. In this limit the excitation-number-non
conserving terms in $H^{(4)}$ are strongly off-resonant and can be
safely neglected:

\begin{align*}
H^{(4)} & \approx-\frac{E_{C}}{4}\left\{ 2b_{0}^{\dagger}b_{0}^{\dagger}b_{0}b_{0}+3b_{1}^{\dagger}b_{1}^{\dagger}b_{1}b_{1}\right.\\
 & \left.+8b_{1}^{\dagger}b_{1}b_{0}^{\dagger}b_{0}+2\left(b_{0}^{\dagger2}b_{1}^{2}+b_{1}^{\dagger2}b_{0}^{2}\right)\right\} 
\end{align*}
where we assumed $L_{x}/\lambda_{J}\gg1$. In conclusion we note that
we find interaction terms corresponding to the self-, cross-Kerr and
parametric interactions between different modes.

\subsection{Interaction with the resonator mode \label{qubit-light coupling}}

We now consider the interaction with the resonator mode assuming the
geometry shown in Fig.~\ref{Fig1}. Ignoring the magnetic effects
we add the coupling in the gauge-invariant way to the $\partial_{\tau}\theta^{\left(-\right)}\rightarrow\partial_{\tau}\theta^{\left(-\right)}-2ie\int_{1}^{2}d\vec{z}\cdot{\bf E}^{\text{res}}$
action Eq.~\eqref{eq:SJ-1}:

\begin{align*}
S_{0} & \approx\int\frac{d^{2}{\bf r}}{S}d\tau\left\{ \frac{1}{16E_{C}}\left(\partial_{\tau}\theta^{\left(-\right)}-2ie\int_{1}^{2}d\vec{z}\cdot{\bf E}^{\text{res}}\right)^{2}+\frac{E_{\text{J}}\lambda_{\text{J}}^{2}}{2}\left(\vec{\nabla}\theta_{{\bf r}}^{\left(-\right)}\right)^{2}\right\} 
\end{align*}
Following the main text we now first reduce the action to the 1-dimensional
form and perform the expansion eigenmode expansion $\theta^{\left(-\right)}\left(x\right)=\sum_{n}\theta_{n}^{(-)}\Xi_{n}\left(x\right)$.
With this we get:

\begin{align*}
S_{0} & \approx\int d\tau\sum_{m}\left\{ \frac{1}{16E_{C}}\left(\partial_{\tau}\theta_{m}^{(-)}\partial_{\tau}\theta_{m}^{(-)}-4ie\partial_{\tau}\theta_{m}^{(-)}\int_{1}^{2}d\vec{z}\cdot{\bf E}_{m}^{\text{res}}-4e^{2}\int\frac{dx}{L_{x}}d\tau\left(\int_{1}^{2}d\vec{z}\cdot{\bf E}^{\text{res}}\right)^{2}\right)\right\} \\
 & +\int\frac{dx}{L_{x}}d\tau\left\{ \frac{E_{\text{J}}\lambda_{\text{J}}^{2}}{2}\left(\vec{\nabla}\theta_{{\bf r}}^{\left(-\right)}\right)^{2}\right\} ,
\end{align*}
where ${\bf E}_{m}^{\text{res}}\equiv\int\frac{dx}{L_{x}}{\bf E}^{\text{res}}\Xi_{m}\left(x\right)$.
We now have to deduce the quantum Hamiltonian of the resonator+junction
system which generates the action above. Due to the linearity, the
easiest way to extract the coupling is to perform Legendre transform:

\[
\pi_{m}^{(-)}=\frac{1}{8E_{C}}\partial_{\tau}\theta_{m}^{(-)}-\frac{1}{4E_{C}}\left(ie\int_{1}^{2}d\vec{z}\cdot{\bf E}_{m}^{\text{res}}\right)
\]
With this we get the coupling Hamiltonian:

\[
\hat{H}_{\text{int}}=2e\int_{1}^{2}d\vec{z}\cdot\hat{{\bf E}}_{m}^{\text{res}}\hat{\pi}_{m}
\]
The Hamiltonian $H_{\text{int}}$ is very general in the sense that it is valid regardless of the details of the system. In realistic scenarios its evaluation represents a complicated problem which requires taking into account e.g. near-field effects (see Sec.~\ref{sec:exp_cons}). In the current section we resort to a simplified parallel-plate geometry. We note that in addition we get terms acting on the electromagnetic
field degrees of freedom only. In the following we assume these terms
are absorbed in the definition of the resonator mode frequency.

Now using the expression for the quantized oscillator momentum and
the resonator mode operator Eq.~\eqref{eq:Ez} we get:

\[
\hat{H}_{\text{int}}\approx2e\delta_{z}\sum_{n}E_{n}^{0}\sqrt{\frac{\omega_{m}}{16E_{C}}}(i\hat{a}_{n}{\cal E}_{n,m}+\text{H.c.})(i\hat{b}_{m}^{\dagger}-i\hat{b}_{m}),
\]
where the interaction form-factors are denoted as ${\cal E}_{n,m}\equiv\int_{-L_{x}/2}^{L_{x}/2}\frac{1}{L_{x}}{\cal E}_{n}\left(x\right)\Xi\left(x\right)$.
In the following we will assume ${\cal E}_{n}\left(x\right)=\sin(\frac{\pi nx}{L_{\text{res}}})$
for $n$ being odd and ${\cal E}_{n}\left(x\right)=\cos(\frac{\pi nx}{L_{\text{res}}})$
for even $n$. In particular, assuming the junction is located exactly
in between the centre of the resonator  we get for several low-energy modes:

\begin{align*}
{\cal E}_{0,0} & =\frac{\sin(\tilde{L})}{\tilde{L}}\\
{\cal E}_{1,1} & =\frac{8\sqrt{2}\pi\tilde{L}\cos\left(2\tilde{L}\right)}{\pi^{2}-16\tilde{L}^{2}}\\
{\cal E}_{2,2} & =\frac{3\sqrt{2}\tilde{L}\sin\left(3\tilde{L}\right)}{\pi^{2}-9\tilde{L}^{2}},
\end{align*}
where $\tilde{L}=L\pi/(2L_{\text{res}})$. With this, we find the coupling
terms $g_{m,n}$ in \eqref{eq:Hcoupl} as 
\[
g_{m,n}=2e\delta_{z}E_{n}^{0}\sqrt{\frac{\omega_{m}}{16E_{C}}}{\cal E}_{m,n}.
\]
It is convenient to represent this expression as a product of an effective
dipole moment of $m-$th mode $d_{m}=2e\delta_{z}{\cal E}_{m,n}$
and a zero-point value of electric field in $n-$th field mode $E_{n}^{0}$.

\section{Derivation of the effective action Eq.~(\ref{eq:S0-1}-\ref{eq:SJ})\label{sec:Derivation-of-the}}

In this appendix we provide the microscopic derivation of the effective
action Eqs.~(\ref{eq:S0}, \ref{eq:SJ-1}).

\subsection{Derivation of the effective action}

We now consider the bilayer action \citet{AS10,SFB20}:

\begin{align*}
S_{0} & =\sum_{i=1,2}\int d^{2}{\bf r}L_{z}d\tau\left\{ \frac{\nu}{2}\left(\partial_{\tau}\theta_{{\bf r}}^{(i)}+e\phi_{{\bf r}}^{\left(i\right)}\right)^{2}+\frac{n_{s}}{2m}\left(\vec{\nabla}\theta_{{\bf r}}^{\left(i\right)}-\frac{e}{c}\vec{A}_{{\bf r}}^{\left(i\right)}\right)^{2}\right\} ,\\
S_{\text{J}} & =-\frac{E_{\text{J}}}{S}\int d^{2}{\bf r}d\tau\cos\left\{ 2\left(\theta_{{\bf r}}^{(1)}-\theta_{{\bf r}}^{(2)}-\frac{e}{c}\int_{1}^{2}d\vec{z}\cdot\vec{A}_{{\bf r},z}\right)\right\} ,
\end{align*}
Let us now transform into the $\pm$ basis by defining new variables
for all the fields $X^{\pm}\equiv X^{(1)}\pm X^{(2)}$:

\begin{align*}
S_{0} & =\sum_{i=\pm}\int d^{2}{\bf r}L_{z}d\tau\left\{ \frac{\nu}{4}\left(\partial_{\tau}\theta_{{\bf r}}^{(i)}+e\phi_{{\bf r}}^{\left(i\right)}\right)^{2}+\frac{n_{s}}{4m}\left(\vec{\nabla}\theta_{{\bf r}}^{\left(i\right)}-\frac{e}{c}\vec{A}_{{\bf r}}^{\left(i\right)}\right)^{2}\right\} 
\end{align*}
The ``-'' variable we redefine in a gauge-invariant way $\theta_{{\bf r}}^{(-)}\rightarrow\theta_{{\bf r}}^{(-)}-\frac{e}{c}\int_{1}^{2}d\vec{z}\cdot\vec{A}_{{\bf r},z}$
by adding and subtracting $\frac{e}{c}\int_{1}^{2}d\vec{z}\cdot\vec{A}_{{\bf r},z}$:

\begin{align*}
S_{0}^{(-)} & =\int d^{2}{\bf r}L_{z}d\tau\frac{\nu}{4}\left(\partial_{\tau}\theta_{{\bf r}}^{(-)}+e\phi_{{\bf r}}^{\left(-\right)}+\frac{e}{c}\int_{1}^{2}d\vec{z}\cdot\partial_{\tau}\vec{A}_{{\bf r},z}\right)^{2}\\
 & +\int d^{2}{\bf r}L_{z}d\tau\frac{n_{s}}{4m}\left(\vec{\nabla}\theta_{{\bf r}}^{\left(-\right)}-\frac{e}{c}\vec{A}_{{\bf r}}^{\left(-\right)}+\frac{e}{c}\int_{1}^{2}d\vec{z}\cdot\nabla_{r}\vec{A}_{{\bf r},z}\right)^{2}
\end{align*}
and the Josephson term reads:

\[
S_{\text{J}}=-\frac{E_{\text{J}}}{S}\int d^{2}{\bf r}d\tau\cos\left\{ 2\theta_{{\bf r}}^{(-)}\right\} ,
\]
We now use the following two relations which are straightforwardly
found using definitions of the scalar and vector potentials: 
\[
e\phi_{{\bf r}}^{\left(i\right)}+\frac{e}{c}\int_{1}^{2}d\vec{z}\cdot\partial_{\tau}\vec{A}_{{\bf r},z}=-ie\int_{1}^{2}d\vec{z}{\bf E}
\]
and 
\[
\frac{e}{c}\int_{1}^{2}dz\left[{\bf B}\times{\bf e}_{z}\right]-\frac{e}{c}\vec{A}_{{\bf r}}^{\left(-\right)}+\frac{e}{c}\int_{1}^{2}d\vec{z}\cdot\nabla_{r}\vec{A}_{{\bf r},z}=\frac{e}{c}\int_{1}^{2}dz\left[{\bf B}\times{\bf e}_{z}\right]
\]
With this we get:

\begin{align*}
S_{0}^{(-)} & =\int d^{2}{\bf r}L_{z}d\tau\left\{ \frac{\nu}{4}\left(\partial_{\tau}\theta_{{\bf r}}^{(-)}-ie\int_{1}^{2}d\vec{z}\cdot{\bf E}\right)^{2}+\frac{n_{s}}{4m}\left(\vec{\nabla}\theta_{{\bf r}}^{\left(-\right)}+\frac{e}{c}\int_{1}^{2}dz\left[{\bf B}\times{\bf e}_{z}\right]\right)^{2}\right\} 
\end{align*}
In order to get the action of the junction alone we now integrate-out
the electromagnetic field by simply solving the Maxwell equations.
We assume that the superconductors are located sufficiently close
such that the retardation can be ignored. Denoting two fields $\int_{1}^{2}dzE_{{\bf r}}^{(z)}$
and $\int_{1}^{2}dz\left[{\bf B}\times{\bf e}_{z}\right]$ we can
find the bare correlation functions according to the Maxwell equations:

\begin{align*}
D_{0,q} & \equiv-\left\langle i\int_{1}^{2}d\vec{z}E_{{\bf q}}^{(z)}i\int_{1}^{2}dzE_{-{\bf q}}^{(z)}\right\rangle \approx2\int_{1}^{2}dz\int_{1}^{2}dz'dk_{z}e^{ik_{z}\left(z-z'\right)}\left(\frac{k_{z}^{2}}{q_{x}^{2}+k_{z}^{2}}\right)\\
 & \approx4\frac{\pi}{q}\left(1-e^{-q\delta_{z}}\right)\approx4\pi\delta_{z}\\
D_{q} & \approx\left\langle \int_{1}^{2}dz\left[{\bf B}_{{\bf q}}\times{\bf e}_{z}\right]_{{\bf q}}\int_{1}^{2}dz\left[{\bf B}_{-{\bf q}}\times{\bf e}_{z}\right]_{-{\bf q}}\right\rangle \approx4\pi\delta_{z}
\end{align*}
The cross-correlation-term yields terms linear in frequency and it
can be ignored at low energies.

\subsubsection{EM field integrated-out}

We now integrate-out the EM field. The resulting action reads:

\[
S_{0}^{\left(-\right)}=\frac{1}{2}\sum_{q,\epsilon_{n}}\left\{ L_{z}\frac{\nu}{2}\epsilon_{n}^{2}\left(\frac{1}{L_{z}\frac{\nu e^{2}}{2}D_{0,q}^{\left(\pm\right)}+1}\right)\theta_{{\bf q},n}^{(-)}\theta_{-{\bf q},-n}^{(-)}+L_{z}\frac{n_{s}}{2m}\left(\frac{1}{L_{z}\frac{n_{s}}{2m}\left(\frac{e}{c}\right)^{2}D_{q}^{\left(\pm\right)}+1}\right){\bf q}^{2}\theta_{{\bf q},n}^{(-)}\theta_{-{\bf q},-n}^{\left(-\right)}\right\} 
\]

The $+$ component is of higher order in space-time derivatives and
we ignore it. We only keep the ``-'' component which at low energy
becomes:

\[
S_{0}^{\left(-\right)}=\frac{1}{2}\sum_{q,\epsilon_{n}}\left\{ \left(\frac{L_{z}\frac{\nu}{2}}{2\pi L_{z}\nu e^{2}\delta_{z}+1}\right)\epsilon_{n}^{2}\theta_{{\bf q},n}^{(-)}\theta_{-{\bf q},-n}^{(-)}+\frac{L_{z}\frac{n_{s}}{2m}}{\frac{L_{z}\delta_{z}}{2\lambda_{L}^{2}}+1}{\bf q}^{2}\theta_{{\bf q},n}^{(-)}\theta_{-{\bf q},-n}^{\left(-\right)}\right\} 
\]
We now denote the charging energy as $E_{C}=\left(\frac{2\pi L_{z}\nu e^{2}\delta_{z}+1}{L_{z}\nu}\right)$
and assume $\frac{L_{z}\delta_{z}}{2\lambda_{L}^{2}}\gg1$ and get:

\begin{align*}
S_{0}^{\left(-\right)} & =\sum_{q,\epsilon_{n}}\frac{1}{S}\left\{ \frac{1}{4E_{C}}\epsilon_{n}^{2}\theta_{{\bf q},n}^{(-)}\theta_{-{\bf q},-n}^{(-)}+2E_{J}\frac{L_{z}\frac{n_{s}}{2m}}{\left(\frac{L_{z}\delta_{z}}{2\lambda_{L}^{2}}+1\right)2\frac{cj_{c}}{2e}}{\bf q}^{2}\theta_{{\bf q},n}^{(-)}\theta_{-{\bf q},-n}^{\left(-\right)}\right\} \\
 & =\sum_{q,\epsilon_{n}}\frac{1}{S}\left\{ \frac{1}{4E_{C}}\epsilon_{n}^{2}\theta_{{\bf q},n}^{(-)}\theta_{-{\bf q},-n}^{(-)}+2E_{J}\lambda_{J}^{2}{\bf q}^{2}\theta_{{\bf q},n}^{(-)}\theta_{-{\bf q},-n}^{\left(-\right)}\right\} 
\end{align*}
with $\lambda_{J}=\sqrt{\frac{c\Phi_{0}}{\left(\delta_{z}+\frac{2\lambda_{L}^{2}}{L_{z}}\right)8\pi^{2}j_{c}}}$.
We note that this expresion for Josephson penetration length is only
valid for thin junctions. We thus find the action Eq.~\eqref{eq:S0},
\eqref{eq:SJ-1} up to a renormalization factor $\theta\rightarrow2\theta$.

\section{Derivation of the Josephson term}

In this section, we derive the Josephson coupling term in the action
Eq.~(\ref{eq:S0-1}-\ref{eq:SJ}). We start with the action describing
two conventional s-wave superconductors with the tunneling between
them characterized by the rate $t$. The imaginary-time action including
only the phase modes reads \citep{AS10}:

\[
S=\int_{0}^{\beta}d\tau\int d^{2}{\bf r}\sum_{n}\Psi_{{\bf r}}^{\dagger}\left(\begin{array}{cccc}
-\partial_{\tau}-\hat{\xi} & \Delta e^{i\theta_{{\bf r}}^{(1)}} & t & 0\\
\Delta e^{-i\theta_{{\bf r}}^{(1)}} & -\partial_{\tau}+\hat{\xi} & 0 & -t\\
t & 0 & -\partial_{\tau}-\hat{\xi} & \Delta e^{i\theta_{{\bf r}}^{(2)}}\\
0 & -t & \Delta e^{-i\theta^{(2)}} & -\partial_{\tau}+\hat{\xi}
\end{array}\right)\Psi_{{\bf r}},
\]
where $\hat{\xi}=-\frac{\nabla^{2}}{2m}-\mu$ is the kinetic energy,
$\Psi=\{\psi_{{\bf r},\uparrow}^{(1)},\psi_{{\bf r},\downarrow}^{\dagger(1)},\psi_{{\bf r},\uparrow}^{(2)},\psi_{{\bf r},\downarrow}^{\dagger(2)}\}^{T}$
is the Nambu spinor describing the electrons electrons and $\Delta$
is the gap. We now perform the gauge transformation \citep{AS10}
and get

\[
S=\int_{0}^{\beta}d\tau\int d^{2}{\bf r}\Psi_{{\bf r}}^{\dagger}\left(\begin{array}{cccc}
-\partial_{\tau}-\hat{\xi} & \Delta & te^{-i\theta_{{\bf r}}^{(-)}} & 0\\
\Delta & -\partial_{\tau}+\hat{\xi} & 0 & -te^{i\theta_{{\bf r}}^{(-)}}\\
te^{i\theta_{{\bf r}}^{(-)}} & 0 & -\partial_{\tau}-\hat{\xi} & \Delta\\
0 & -te^{i\theta_{{\bf r}}^{(-)}} & \Delta & -\partial_{\tau}+\hat{\xi}
\end{array}\right)\Psi_{{\bf r}},
\]
where $\theta_{{\bf r}}^{(-)}$ is the phase difference. Let us now
integrate-out the electron gas to one loop assuming the tunneling
is weak. We get in the limit of low momenta:

\begin{align*}
S_{{\rm J}}^{(2)} & =t^{2}\int d^{2}{\bf r}\sum_{m,n}F_{m}\left\{ \cos(\theta_{r}^{(-)})\right\} _{n}\left\{ \cos(\theta_{{\bf r}}^{(-)})\right\} _{m-n}\\
 & +t^{2}\int d^{2}{\bf r}\sum_{m,n}B_{m}\left\{ \sin(\theta_{r}^{(-)})\right\} _{n}\left\{ \sin(\theta_{{\bf r}}^{(-)})\right\} _{m-n},
\end{align*}
where the subscripts indicate taking  the corresponding bosonic
discrete temporal Fourier transform component $\Omega_{n}=\frac{2\pi n}{\beta}$,
$\Omega_{m}=\frac{2\pi m}{\beta}$,

\begin{align}
F_{m} & =\frac{1}{\beta V}\sum_{{\bf k},l}\text{Tr}\left[\hat{G}_{{\bf k}}^{(0)}(i\epsilon_{l}-i\Omega_{m})\hat{\tau}_{3}\hat{G}_{{\bf k}}^{(0)}(i\epsilon_{l})\hat{\tau}_{3}\right]\label{eq:A_m}\\
B_{m} & =\frac{1}{\beta V}\sum_{{\bf k},l}\text{Tr}\left[\hat{G}_{{\bf k}}^{(0)}(i\epsilon_{l}-i\Omega_{m})\hat{\tau}_{0}\hat{G}_{{\bf k}}^{(0)}(i\epsilon_{l})\hat{\tau}_{0}\right],\label{eq:B_m}
\end{align}
and $\epsilon_{l}=\pi\left(2l+1\right)/\beta$ is the Fermionic Matsubara
frequency, $\hat{G}_{k}^{(0)}(i\epsilon_{l})=\left(i\epsilon_{l}\hat{\tau}_{0}-\xi_{k}\hat{\tau}_{3}-\Delta\hat{\tau}_{1}\right)^{-1}$
is the unperturbed mean-field Green's function of either superconductors
without tunneling, $\hat{\tau}_{i}$ is $i$-th Pauli matrix. Upon
taking integrals in Eqs.~(\ref{eq:A_m}, \ref{eq:B_m}) we find $B_{m}=0$
and

\[
F_{m}=-\frac{\nu_{0}}{\pi}\frac{4\Delta\text{arccosh}(\frac{\sqrt{4\Delta^{2}+\Omega_{m}^{2}}}{2\Delta})}{\left|\Omega_{m}\right|\sqrt{4\Delta^{2}+\Omega_{m}^{2}}},
\]
where $\nu_{0}$ is the normal-state density of states. In the limit
of low energies $\Omega_{m}\ll\Delta$ we find $F_{m}\approx-\frac{\nu_{0}}{\pi}(1+\frac{\Omega_{m}^{2}}{6\Delta^{2}})$.
Ignoring the $\frac{\Omega_{m}^{2}}{6\Delta^{2}}$ term in the Markov-like
approximation we find the Josephson term:

\begin{align*}
S_{{\rm J}}^{(2)} & \approx-\frac{t^{2}\nu_{0}}{\pi}\int d^{2}{\bf r}\sum_{n}\left\{ \cos(\theta_{r}^{(-)})\right\} _{n}\left\{ \cos(\theta_{{\bf r}}^{(-)})\right\} _{-n}\\
 & =-\frac{t^{2}\nu_{0}}{\pi}\int d^{2}{\bf r}\int d\tau\left\{ \cos(\theta_{r}^{(-)})\right\} ^{2}\\
 & \thickapprox-\frac{t^{2}\nu_{0}}{2\pi}\int d^{2}{\bf r}\int d\tau\cos(2\theta_{r}^{(-)})
\end{align*}
In conclusion we note that we neglected the finite-momentum contributions
to $B$ which are of the order of $k/(v_{F}/\Delta)$ where $k$ is
the characteristic momentum of the phase fluctuations.

\section{Nucleation of Solitons\label{sec:Nucleation-of-Solitons}}

We now study the possibility spontaneous nucleation of sine-Gordon
solitons. For a sufficiently large junction the estimate of the soliton
nucleation rate $\sim e^{-\beta E_{0}}$ can be obtained based on
the energy $E_{0}$ required for a creation of of a kink-antikink
\citep{KM89} to be determined below. The derivation is rather tedious
but all the necessary energy and space scales can be understood in
the classical limit. To this end, we now write the quantum sine-Gordon
action in the following form:

\begin{equation}
\frac{S}{\hbar}=\frac{1}{8\hbar E_{C}}\int_{0}^{\beta\hbar}d\tau\frac{dx}{L_{x}}\left(\frac{1}{2}\left(\partial_{\tau}\theta\right)^{2}+\frac{\omega_{pl}^{2}\lambda_{J}^{2}}{2}\left(\partial_{x}\theta\right)^{2}-\omega_{pl}^{2}\cos\left(\theta\right)\right),\label{eq:Lsg-2}
\end{equation}
where we kept explicitly the $\hbar$. It is now convenient to rescale
the imaginary time as $\tau\rightarrow\hbar\beta\tau$. In this case
we find:

\begin{equation}
\frac{S}{\hbar}=\frac{\beta}{8E_{C}}\int_{0}^{1}d\tau\frac{dx}{L_{x}}\left(\frac{1}{2\beta^{2}\hbar^{2}}\left(\partial_{\tau}\theta\right)^{2}+\frac{\omega_{pl}^{2}\lambda_{J}^{2}}{2}\left(\partial_{x}\theta\right)^{2}-\omega_{pl}^{2}\cos\left(\theta\right)\right),\label{eq:Lsg-2-1}
\end{equation}
by taking the limit of $\hbar\rightarrow0$ we find that the time
variations of the phase produce very large Boltzmann weight. By neglecting
them we find the action describing classical (thermal) nucleation
of solitons:

\begin{equation}
S_{cl}=\lim_{\hbar\rightarrow0}\frac{1}{\hbar}S_{cl}=\frac{1}{8E_{C}}\beta\int\frac{dx}{L_{x}}\left(\frac{\omega_{pl}^{2}\lambda_{J}^{2}}{2}\left(\partial_{x}\theta\right)^{2}-\omega_{pl}^{2}\cos\left(\theta\right)\right),\label{eq:Lsg-2-1-1}
\end{equation}
The relevant length and energy scales are thus given by $\lambda_{J}$
and $E_{J}=\omega_{pl}^{2}/E_{C}$ respectively. Precise estimate
of the number of solitons per unit length in equilibrium is given
by \citep{W80,KM89}:

\[
\left\langle n\right\rangle =\left(\frac{2}{\pi}\right)^{1/2}\lambda_{J}^{-1}\sqrt{\frac{2E_{J}}{k_{{\rm B}}T}}e^{-2E_{J}/k_{{\rm B}}T}
\]
Using the temperature estimate from \citet{MHC21}, $T=20$mK we find
$E_{J}/k_{{\rm B}}T\sim80$ and thus for the junction sizes of the
order of $L\sim\lambda_{J}$ the nucleation can be safely neglected.

\section{Experimental characterization proposal\label{subsec:Experimental-characterization-pr}}

We now consider how the nonlinear terms in Eq.~\eqref{eq:H4-1} can
be probed in an experimental setting. For simplicity we only focus
on the Kerr and cross-Kerr terms. The setup we have in mind is shown
in Fig.~\ref{Fig2}~(a). We assume that the junction is placed precisely
in the center of the resonator and therefore each of the lowest modes
can be addressed individually as discussed in Sec.~\ref{subsec:Coupling-to-the}.
The resonator is driven by a two microwave tones, nearly resonant with
the lowest junction modes. In the far-detuned regime, i.e. when the driving frequency is significantly off-resonant with the resonator,
we can assume the resonator field to be classical. In this case the
external diving can be described by the RWA Hamiltonian $H_{\text{dr}}(t)=\sum_{n=1,2}\alpha_{n}\left(\hat{b}_{n}e^{i\omega_{n}^{\text{dr}}t}+\hat{b}_{n}^{\dagger}e^{-i\omega_{n}^{\text{dr}}t}\right)$,
where $\alpha_{n}$ is the slowly varying amplitude of the $n$-th
resonator mode and we impose $\omega_{n}^{\text{dr}}\approx\omega_{n}$.
Moreover we assume the following condition for the driving strengths
$\alpha_{0}\gg\alpha_{1}$. Neglecting the rapidly rotating terms
the set of Heisenberg equations of motion in the mean-field approximation
becomes:

\begin{align}
\frac{d}{dt}\hat{b}_{0} & \approx-i\left(\Delta_{0}-E_{C}\chi_{0,0}|\langle\hat{b}_{0}\rangle|^{2}\right)\hat{b}_{0}-i\alpha_{0},\label{eq:dtb0}\\
\frac{d}{dt}\hat{b}_{1} & \approx-i\left(\Delta_{1}-\frac{E_{C}}{3}\chi_{0,1}|\langle\hat{b}_{0}\rangle|^{2}\right)\hat{b}_{1}-i\alpha_{1},\label{eq:dtb1}
\end{align}
where the detuning is denoted as $\Delta_{n}=\omega_{n}-\omega_{n}^{\text{dr}}$.
We note that here we assumed that the decoherence is neglected for
simplicity. Its effect can be taken into account by introducing the
decoherence terms to the righthand sides of Eqs.~(\ref{eq:dtb0}-\ref{eq:dtb1})
as well as the appropriate Langevin noise terms.

From Eqs.~(\ref{eq:dtb0}-\ref{eq:dtb1}) we find that the resonance
frequency of both junction modes, given the term in parentheses, depends on
amplitude of $0$-th junction mode. By changing the driving intensity the resonances
will be shifted as shown schematically in Fig.~\ref{Fig2}~(b).
This can be probed by measuring the resonator transmission which directly
reflects the amplitude of each mode \citep{SZ99}. We note that in
the discussion above we neglected the possible decoherence effects
in the junction.

\begin{figure}[h]
\begin{centering}
\includegraphics[scale=0.4]{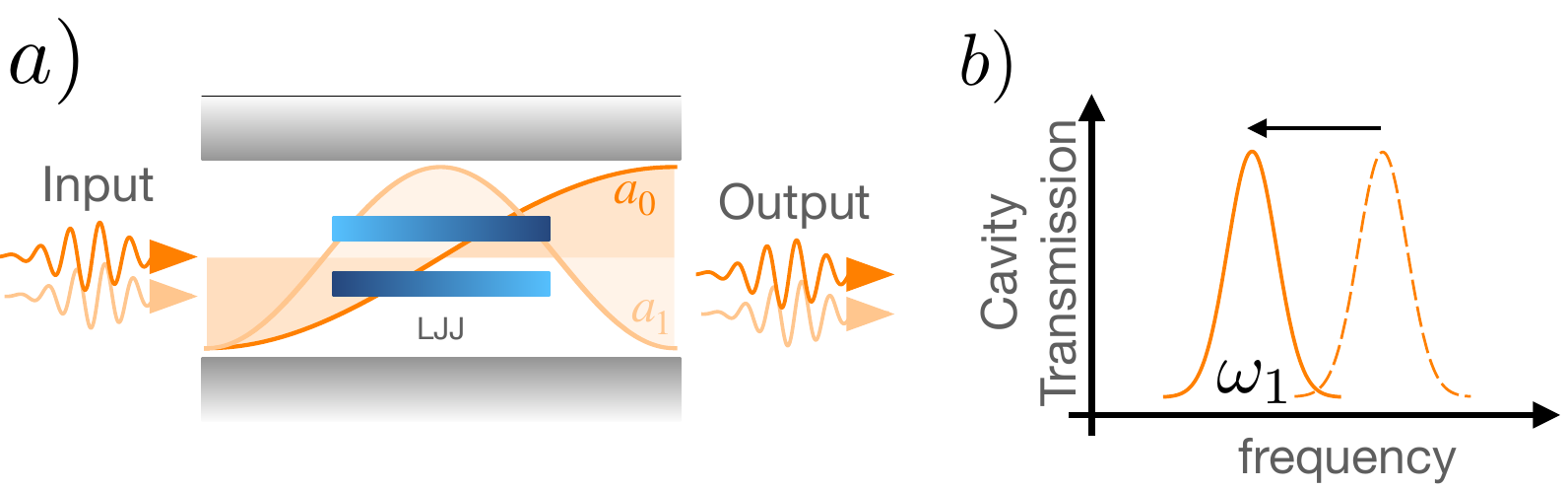} 
\par\end{centering}
\caption{Figure cross-Kerr effect characterization. (a) Scheme of a setup for
probing the Kerr and cross-Kerr effects. (b) Resonant driving of one
of the junction eigenmodes leads to a frequency shift of another mode
which can be probed by a weak laser.}

\label{Fig2} 
\end{figure}
\end{widetext}

\end{document}